\documentstyle[psfig]{l-aa}
\def\jref#1 #2 #3 #4 {{\par\noindent \hangindent=2em \hangafter=1
      \advance \rightskip by 0em #1, {\it#2}, {\bf#3}, #4.\par}}
\def\rref#1{{\par\noindent \hangindent=2em \hangafter=1
      \advance \rightskip by 0em #1.\par}}

\def\COBE{{\sl COBE\/}}
\def\wisk#1{\ifmmode{#1}\else{$#1$}\fi}

\def\etal   {et~al.\,}

\def\deg{\ifmmode^{\circ}\else$^{\circ}$\fi} 
\def\min{\ifmmode^{\prime}\;\else$^{\prime}\;$\fi}
\def\sec{\ifmmode^{\prime\prime}\;\else$^{\prime\prime}\;$\fi}

\def\dn{\ifmmode{\Delta\nu{_d} }\else{$\Delta\nu_{d}$ }\fi}
\def\dt{\ifmmode{\Delta t{_d} }\else{$\Delta t_{d}$ }\fi}

\def\lsim{\,\lower2truept\hbox{${< \atop\hbox{\raise4truept\hbox{$\sim$}}}$}\,}
\def\gsim{\,\lower2truept\hbox{${> \atop\hbox{\raise4truept\hbox{$\sim$}}}$}\,}

\title{{\sc Planck} LFI: Comparison Between Galaxy Straylight Contamination and other
Systematic Effects}

\author{{C. Burigana}\inst{1} 
\and {D. Maino}\inst{2} 
\and {K.M. G\'orski}\inst{3,4}
\and {N. Mandolesi}\inst{1} 
\and {M. Bersanelli}\inst{5} 
\and {F. Villa}\inst{1} 
\and {L. Valenziano}\inst{1} 
\and {B.D. Wandelt}\inst{6}
\and {M. Maltoni}\inst{7} 
\and {E. Hivon}\inst{8}
} 
\begin{document}
\offprints{burigana@tesre.bo.cnr.it}
\date{Submitted on A\&A, 28.9.2000}
\thesaurus{12(12.03.1; 12.04.2) 10(10.07.1) 03(03.19.2; 03.20.9; 03.13.2)}
\institute{
{Istituto TeSRE, Consiglio Nazionale delle Ricerche, via Gobetti 101, I-40129 Bologna, Italy} 
\and
{Oss. Astr. Trieste, via G.B.~Tiepolo 11, I-34131 Trieste, Italy}
\and
{ESO, European Southern Observatory, Karl-Schwarzschild Str. 2, D-85748 Garching, Germany}
\and
{Warsaw University Observatory, Warsaw, Poland}
\and
{Dipartimento di Fisica, Universit\`a di Milano, and IFC/CNR, via Celoria 16, I-20133, Milano, Italy} 
\and
{Department of Physics, Princeton University, Princeton, NJ-08544, USA}
\and
{Instituto de F\'{\i}sica Corpuscular~--~CSIC/UVEG, Edificio Institutos de Paterna, Apt.~22085, E-46071 Valencia, Spain}
\and
{Observational Cosmology, California Institute of Technology, MS 59-33, CA-91125 Pasadena, U.S.A.}
}

\maketitle
\markboth{C. Burigana et al.: Galaxy straylight
contamination versus other effects in LFI observations}
{C. Burigana et al.: Galaxy straylight
contamination versus other effects in LFI observations}

\begin{abstract}
The straylight contamination due to the Galactic emission 
(GSC, Galaxy Straylight Contamination)
entering at large angles from the antenna centre direction  
may be one of the 
most critical sources of systematic effects in observations
of the cosmic microwave background (CMB) anisotropies
by future satellite missions like {\sc Planck} and MAP.
For the Low Frequency Instrument (LFI), 
this effect is expected to be particularly crucial
at the lowest frequency channels.
We describe here different methods to evaluate the impact of this effect
and compare it with other systematics of instrumental and astrophysical origin.
The results are presented in terms of simulated data streams and 
maps, Fourier series decomposition and angular power spectrum.
The contributions within
few degrees from the beam centre dominate 
the GSC near the Galaxy plane.
The antenna sidelobes at intermediate and large angles 
from the beam centre
dominate the GSC at medium and high Galactic latitudes. 
We find a GSC peak at $\sim 15\mu$K and a GSC
angular power spectrum above that of the white noise for multipoles 
$\ell \lsim 10$ albeit smaller than that of CMB anisotropies by a factor 
larger than $\sim 10$.
At large multipoles,  
the GSC affects the determination of CMB angular power spectrum 
significantly less than other kinds of instrumental 
systematics, like main beam distortions and $1/f$ noise.
Although the GSC is largest at low 
Galactic latitudes, the contamination
produced by far pattern features at medium and
high Galactic latitudes, peaking at $\sim 4\mu$K,  
has to be carefully investigated, 
because the combination of low amplitude of Galaxy emission in those regions
with the extremely good nominal {\sc Planck} sensitivity 
naturally makes high Galactic latitude areas the targets for
unprecedentedly precise estimation of cosmological CMB anisotropy.
This paper is based on {\sc Planck} LFI activities.

\keywords{\it Cosmology: cosmic microwave background~-- Galaxy: general -- Space vehicles --
Telescopes~-- {Methods}: data analysis}

\end{abstract}

\section{Introduction}

After the great success of \COBE-DMR 
(Smoot \etal 1992, Bennet \etal 1996a, G\'orski \etal 1996)
which probed the gravitational instability scenario 
for structure formation through the detection of CMB anisotropies at few degree scales,
and the recent balloon-borne experiments 
at high sensitivity and 
resolution on limited sky regions 
(De~Bernardis \etal 2000, Hanany \etal 2000), 
supporting a universe model with $\Omega_{tot} \sim 1$ 
(Lange \etal 2000, Balbi \etal 2000, Jaffe \etal 2000), 
ultimately, future progresses of the CMB anisotropy cosmology will be based on
two space missions, MAP (Microwave Anisotropy Probe) 
(see Bennet \etal 1996b) by NASA 
and {\sc Planck}
by ESA, planned to be launched respectively in the years
2001 and 2007.

In particular, the Low Frequency Instrument (LFI, Mandolesi \etal 1998) and 
the High Frequency Instrument (HFI, Puget \etal 1998) on-board {\sc Planck} 
will cover together a wide frequency range (30$\div$900 GHz) which 
should significantly improve the accuracy of the subtraction 
of foreground contamination from the primordial CMB anisotropy, 
providing at the same time a gold mine
of cosmological as well as astrophysical information (e.g. De~Zotti \etal 1999a
and references therein).

To fully reach these scientific goals,
great attention has to be devoted 
to properly reduce and/or subtract all the possible systematic effects.
Detailed simulation codes have been developed and are continuously
implemented to analyse the impact of several classes of instrumental effects
related to the
behaviour of the optics, instruments and environment 
for a wide set of possible scanning strategies
(e.g. Burigana \etal 1998, Delabrouille 1998, Maino \etal 1999,
Mandolesi \etal 2000a). Ultimately this effort leads to the optimization
of the mission design and the production of realistic 
data streams and simulated maps for data analysis tools as well as 
a realistic definition of {\sc Planck} scientific performance.

In particular, 
the behaviour of the {\sc Planck} antenna patterns both at intermediate
and large angles from the directions of beam centres have to be carefully considered.
The requirement on the rejection of unwanted radiation coming from directions far
from the optical axis (straylight) is stringent for {\sc Planck} and does not pertain only
the telescope itself, but the
entire optical system, including also solar panels, shielding, 
thermal stability and focal assembly components.
The primary sources of error for the LFI are those due to
imperfect off-axis rejection by the optical system of radiation 
from the Sun, Earth, Moon, planets, Galaxy and the spacecraft itself 
(de~Maagt \etal 1998).
The variations of the spurious straylight signal during the mission 
introduce contaminations in the anisotropy measurements.

The antenna response features at large angular scales from the beam centre 
(far sidelobes) are determined largely by
diffraction and scattering from the edges of the mirrors and from nearby
supporting structures. Therefore they can be reduced by decreasing the illumination at
the edge of the primary, i.e. increasing the edge taper, 
defined as the ratio of the power per unit area
incident on the centre of the mirror to that incident on the edge.
Of course, the higher is edge taper, the lower is the sidelobe level
and the straylight contamination.
On the other hand, increasing the edge taper
has a negative impact on the angular resolution 
for the fixed size of the primary mirror (e.g. Mandolesi \etal 2000a). 
A trade off between angular resolution and straylight contamination 
has to be found.

The main aim of this work is to evaluate the impact of 
the Galactic emission as a source of straylight for {\sc Planck} LFI.
We will then compare it with the effects generated by 
other kinds of systematics, 
the main beam distortion introduced by optical aberrations and the $1/f$ noise 
related to gain fluctuations in LFI radiometers, 
and with the astrophysical contamination from 
the Galaxy and the extragalactic sources in the main beam.
At LFI frequencies, 
the Galaxy straylight contamination (GSC)
is expected to be particularly crucial
at the lowest frequencies, due to the increase of synchrotron absolute
emission and anisotropies with the wavelength. 
For simplicity, we limit our analysis 
to the case of the 30~GHz channel, but the methods presented here
can be extended to higher frequencies.

In Sect.~\ref{simul} we briefly describe the basic recipes of our simulation
code, discussing the geometrical aspects relevant for the full sky convolution,
the format of the optical simulation output performed 
by the ESA (de~Maagt et~al.~1998) that are adopted in the present work,
the conversion from data streams to maps and the computation
of the Fourier modes and of the angular power spectra, two different 
estimators of the GSC impact.
We estimated the expected GSC on the basis of 
the antenna integrated response
from angular regions at different angles from the beam centre, 
and on the level of Galaxy emission.
In Sect.~\ref{acc_test} we focus on the integration accuracy
of our computations
and test the consistency of the code 
by assuming simplified input maps and antenna patterns.
The main results concerning the evaluation of the GSC are presented
in Sect.~\ref{results}. In Sect.~\ref{comparison} they are compared 
with the effects introduced by other kinds of instrumental effects 
and several sources of astrophysical contamination.
Finally, we draw our main conclusions in Sect.~\ref{conclusions}.

\section{Simulation of {\sc Planck} observations}
\label{simul}

The selected orbit for {\sc Planck} is a Lissajous orbit 
around the Lagrangian point L2 of the Sun-Earth system (e.g. Mandolesi \etal 1998). 
The spacecraft spins at 1 r.p.m. 
and the field of view of the two instruments is $5^\circ$ around the
telescope optical 
axis at a given angle $\alpha$ from the spin-axis direction,
given by a unit vector, $\vec s$, choosen outward the Sun direction.
In this work we consider values of  $\alpha \sim 80^{\circ}-90^{\circ}$
\footnote{Most recently it was recommended to choose $\alpha = 85^{\circ}$.}.
The spin axis will be kept parallel to the Sun--spacecraft direction
and repointed by $\simeq 2.5'$ every $\simeq 1$~hour.
In addition, a precession
of the spin-axis with a period, $P$, of $\simeq 6$~months at
a given angle $\beta \sim 10^{\circ}$
about an axis, $\vec f$, parallel to the Sun--spacecraft direction
(and outward the Sun) and shifted of $\simeq 2.5'$ every $\simeq 1$~hour,
may be included in the effective scanning strategy.
These kinds of scanning strategies do not modify the
angle between the spin axis and the Sun--spacecraft direction, avoiding
possible thermal fluctuations induced by modulations 
of the Sun illumination, and allow to achieve nearly full or 
full sky coverage.
Hence {\sc Planck} will trace large circles in the sky.
The detailed distribution in the sky of the number of observations per pixel
depends on the adopted scanning strategy, the telescope design
and the arrangement of the feed array on the telescope focal surface.
The scanning strategy and spacecraft geometry 
have to be carefully addressed in order to minimize 
systematic effects before and after the data analysis
and for ensuring that the sky coverage
be as complete and spatially smooth as possible.

The code we have implemented for simulating 
{\sc Planck} observations for a wide set of scanning 
strategies is described in detail in Burigana \etal (1997, 1998) 
and in Maino \etal (1999).
Here we consider simple scanning strategies, namely with the spin axis
kept always in the ecliptic plane. Under this assumption,
the geometrical input parameters relevant for the scanning strategy are 
the angle $\alpha$ between the spin axis and the telescope line of sight
and the beam location in the telescope field of view.

\noindent
In this study we neglect the small effects introduced on the GSC evaluation 
by the {\sc Planck} orbit by simply assuming {\sc Planck} located in L2
and consider spin-axis shifts of 2\deg every two days
and 180 samplings per scan circle. 
These simplifications allow to speed up the simulation without 
significantly affecting 
our understanding of the main effects introduced by the GSC, 
because of the decreasing of the 
Galaxy fluctuation level at small scales and 
because the effects of pattern features we want to study here occur 
at $\sim$ degree or larger scales. In fact, a recent study 
of synchrotron emission (Baccigalupi et al. 2000a) based on recent 
high resolution surveys at low at medium latitudes indicates 
a steeper power law of the total intensity angular power spectrum where
diffuse emission dominates.

\subsection{Coordinate transformations and input antenna pattern}
\label{coord}
For the simple scanning strategies considered here, 
the vector $\vec s$, which gives the spin axis direction, 
and a unit vector $\vec p$,  parallel to the telescope line of sight $z$,
are simple functions of the time and of three scanning strategy parameters:
the spinning frequency $f_s$, the angle $\alpha$ 
and the orientation of the {\sc
Planck} telescope line of sight at the beginning of each scan circle 
that we arbitrarily fix
close as much as possible to the positive $z$ ecliptic axis. 
In additon, we set
the rotation of $\vec p$ clockwise with respect to the positive direction of $\vec s$.

On the plane tangent to the celestial
sphere in the central direction of the field of view,
i.e. on the field of view plane of the {\sc Planck} telescope,
we choose two coordinates $x$ and $y$, with unit vector 
$\vec u$ and $\vec v$ respectively,
according to the convention that the unit vector 
$\vec u$ points always toward $\vec s$
and that $x,y,z$ is a standard cartesian frame, 
referred here as ``telescope frame''
\footnote{We note that the ``telescope frame'' defined above may be or not
equivalent to the ``primary mirror frame'' depending on the absence or presence
of a tilt angle between the primary mirror axis and the central direction
of the sky field of view. For example, the current optical configuration
forseen for {\sc Planck} exibits a tilt angle of $\simeq - 3.75^{\circ}$.}.

In general, the beam centre
will be identified by its unit vector $\vec b$ in the
frame $x,y,z$ or equivalently by
the coordinates, $x_0,y_0$, of its projection on the plane $x,y$
or, more usually, by its corresponding standard polar coordinates,
the colatitude $\theta_B$ and the longitude $\phi_B$.
The HFI feedhorns will be located in a circular area at the
centre of the focal plane, and LFI feedhorns in a ring
around HFI. Therefore the corresponding positions
of LFI beams on the sky field of view
are significantly off-axis.
For a telescope with $\simeq$~1.5~m aperture the
typical 100~GHz LFI beam is located at $\simeq 2.8^{\circ}$
from the optical axis, whereas the 30~GHz beams are at about
$\simeq 5^{\circ}$ from it.

The shape of the main beam computed by ESA 
(de Maagt \etal 1998) is provided 
in a regular equispaced grid on $x,y$ about the beam centre. 
We can then perform the convolution of the main beam 
(within a choosen angle from the beam centre) with the sky signal 
directly in this frame.

According to the standard output of the ``GRASP8'' code for optical 
simulations as performed de Maagt et~al. (1998), 
we describe the antenna pattern response, $J$, at large angles from the beam centre
by using two standard polar coordinates $\theta_{bf}$ (between
$0^{\circ}$ and $180^{\circ}$)  and $\phi_{bf}$ (between $0^{\circ}$ and $360^{\circ}$)
referred to the ``beam frame''. This 
corresponds to the standard cartesian ``beam frame'' $x_{bf},y_{bf},z_{bf}$ which is obtained by
the ``telescope frame'' $x,y,z$ when the unit vector of the axis $z$ is rotated by an angle
$\theta_B$ on the plane defined by the unit vector of the axis $z$ and the unit vector $\vec b$ up
to reach $\vec b$ \footnote{Properly, the standard of ``GRASP8'' code for optical simulations
is to provide the far pattern in terms of a ``colatitude'' angle $\theta'_{bf}$
(between $-180^{\circ}$ and $180^{\circ}$)  from the ``polar'' axis parallel
$z_{bf}$ and an ``azimuthal'' angle $\phi'_{bf}$ (between $0^{\circ}$ and
$180^{\circ}$) related to $\theta_{bf}$ and $\phi_{bf}$ by:
$\phi'_{bf}=\phi_{bf}$ if $\phi_{bf} \le 180^{\circ}$ and
$\phi'_{bf}=\phi_{bf}-180^{\circ}$ if $\phi_{bf} > 180^{\circ}$;  $\theta'_{bf}=\theta_{bf}$ if
$\phi_{bf} \le 180^{\circ}$ and $\theta'_{bf}=-\theta_{bf}$ if $\phi_{bf} > 180^{\circ}$.}.
We use here the polar coordinates $\theta_{bf},\phi_{bf}$ 
for the convolution of the antenna pattern with the sky signal at large
angles from the beam centre.
For the antenna pattern at ``intermediate'' 
(namely up to few degrees from the beam centre) and ``far'' 
(namely for the entire solid angle) angular distances from the beam centre
different equispaced grids in
$\theta_{bf}$ are available: more
refined for the former, because the response variations are  
stronger close to the beam centre, and less refined for the latter, 
where the relevant response variations occur on degree or larger 
angular scales. 

The orientation of these frames as the satellite moves is implemented in 
the code. For each integration time, we determine the orientations
in the sky of the telescope frame and of the beam frame
and compute the 
pattern response in each considered sky direction, 
thus performing a direct convolution with the sky signal. 

\subsection{Optical results} 
\label{optic}
\begin{figure}[tb]
\vskip 1truecm
\centerline{
\psfig{figure=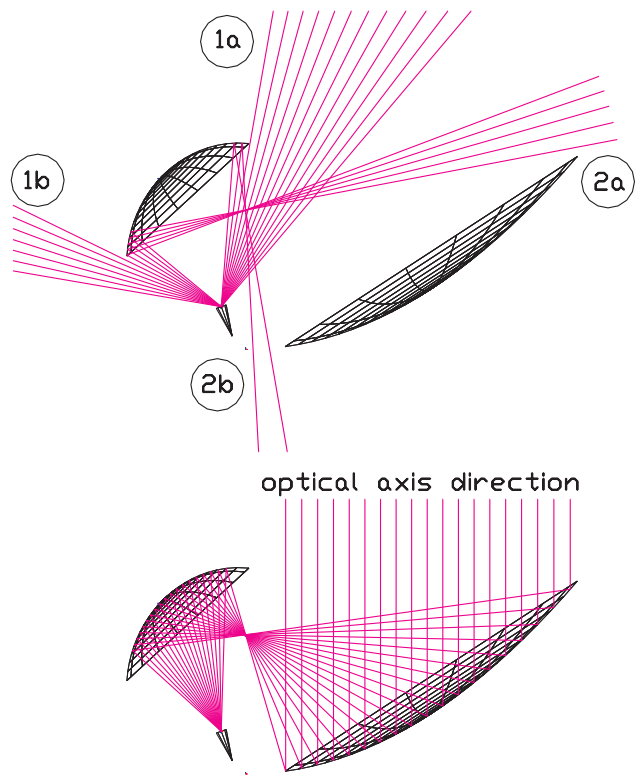,width=8.5cm,height=12.02cm,angle=0}}
\caption[]{Sketch of the telescope design in the symmetry plane ($
\phi'_{bf}\sim 0$). This figure shows the carrier
telescope configuration: a gregorian Dragone-Mizuguchi design with the
primary mirror enlarged to 1.5 meter for reducing the main spillover (2a).
The upper panel shows the rays coming from unwanted directions and focussed
into the feed: (1a) spillover past the subreflector; (1b) spillover past the
subreflector redirect by the shields; (2a) spillover past the main
reflector; (2b) spillover past the main reflector and redirect by the
shields.
The bottom drawing shows the telescope with rays coming from the direction
of the optical axis and focussed in a feed located at the centre of the
focal plane. 
The spillover blocked by the shield, (1b) and (2b), is redirected close to the
main beam as showed on the full-antenna pattern in Fig.~\ref{full_patt_fig}.}
\label{sketch_tel}
\end{figure}

\begin{figure}[tb]
\vskip 1truecm
\centerline{
\psfig{figure=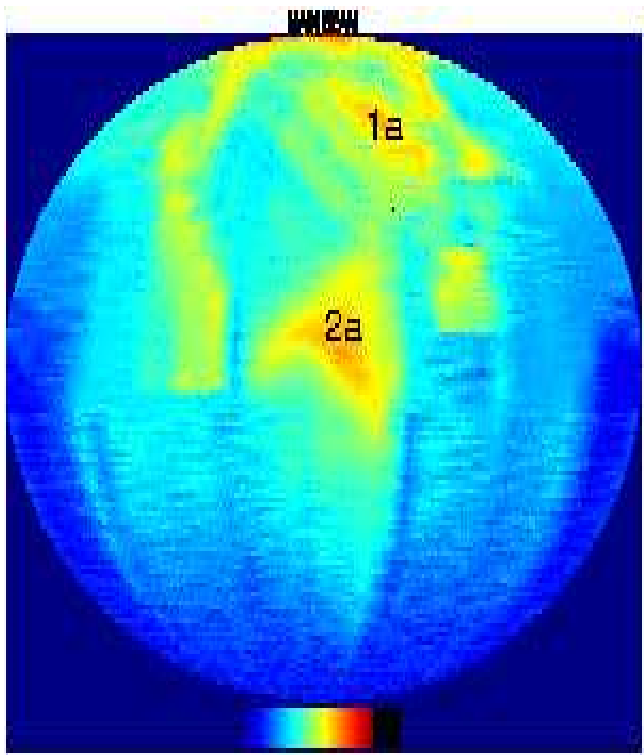,width=8.5cm,height=10cm,angle=-180}}
\caption[]{Full antenna pattern, with response normalized to the maximum, 
for the carrier configuration. The color table is linear
in dB and the true directivity at the maximum is 49.36~dB. Pattern regions
related to particular optical structures are identified. See also the text.}
\label{full_patt_fig}
\end{figure}

\begin{figure}[tb]
\vskip 1truecm
\centerline{
\psfig{figure=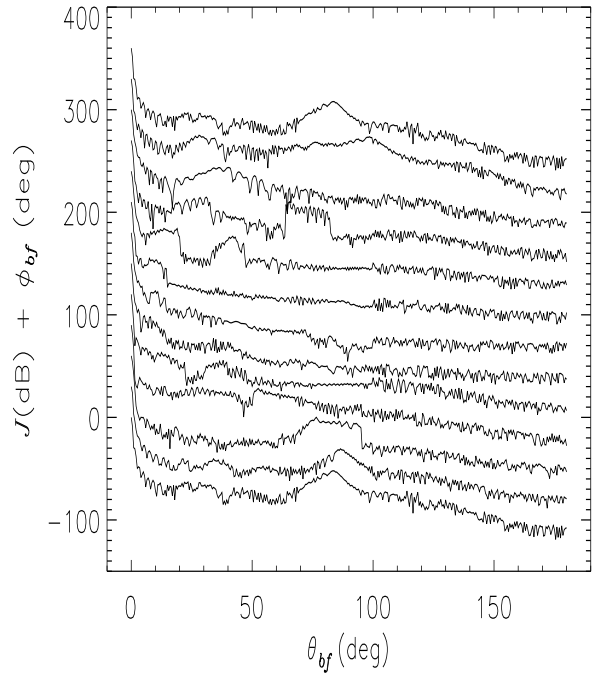,width=8.5cm,height=10cm,angle=0}}
\caption[]{Cuts of full antenna pattern in dB for the carrier configuration.
The lines refer to the antenna response for $\theta_{bf}$
between $0^{\circ}$ and $180^{\circ}$ and, from the bottom to the top, with
$\phi_{bf}$ at steps of $30^{\circ}$ from $0^{\circ}$ to $360^{\circ}$
(each cut is vertically
shifted for graphic purposes to have the pattern response value at 
$\theta_{bf}=0^{\circ}$ equal to the considered value of $\phi_{bf}$).}
\label{full_patt_cuts}
\end{figure}

The telescope design of the Carrier Configuration for {\sc Planck} 
-- the current baseline  in which
{\sc Planck} and FIRST are lauched together and separate in orbit --
is based on the Phase A study. The configuration is a off-axis
Gregorian Dragone-Mizuguchi telescope with the main reflector oversized at
1.5 meter projected aperture for reducing the spillover at the primary mirror. The
subreflector axis is tilted at $14^\circ$ with respect to the main
reflector axis.
The off-axis design of {\sc Planck} telescope (see
Fig.~\ref{sketch_tel}) introduces particular features in the
full sky antenna response. In Fig.\ref{sketch_tel} a sketch on the
symmetry plane of
the telescope is reported. For simplicity, a feedhorn is located at
the centre of the focal plane. Its pattern on the sky is along the
direction of the optical axis of the telescope (see bottom panel of
Fig. \ref{sketch_tel}). 
The unwanted ray directions are shown in the upper panel of
Fig.\ref{sketch_tel}. The rays labelled with 1 are coming from
the sky directly into the feed. The rays labelled with 2 are those
scattered into the feed by the subreflector only.
These identified regions are the spillover past the subreflector for rays
1 and the spillover past the main reflector for rays 2.
Considering the {\sc Planck} Carrier Configuration (de Maagt \etal 1998), 
the rays 1b and 2b are
blocked by the spacecraft shields, which redirect the rays in an angular region close
to the main beam. 
In Fig.~\ref{full_patt_fig} we show a map of the considered full pattern;
see the connection beetween the regions marked in Fig.~\ref{sketch_tel}
and Fig.~\ref{full_patt_fig}.
In Fig.~\ref{full_patt_cuts} we show also
several cuts at constant azimuths $\phi_{bf}$ from 0\deg to 360\deg 
(from the bottom to the top). 


We have conservatively considered the worst case (de Maagt \etal 1998)
for what concerns the straylight effect:
we use the antenna pattern
computed at 30 GHz, the channel with the highest spillover and with the
highest Galaxy signal. We also included the shields for the Carrier
Configuration.

The main feature is the spillover (2a) at about $90^\circ$ from the
main beam (see also Fig.~\ref{full_patt_fig},
where the main beam is located at North pole)
which shows
a response of $\approx -60$~dB with respect to the maximum  
extending for few tens of degrees in 
$\theta_{bf}$ and in $\phi_{bf}$ (around
$\phi_{bf}=0^\circ$, i.e. always quite close to the direction 
of the axis $x$ in the ``telescope frame''). 
Another relevant
feature is the subreflector spillover (1a), with similar response
level and an angular extension close to the main beam ($\theta_{bf}\sim
30^\circ$) as we can see also in Fig.\ref{sketch_tel}. 
Other features are located on the northern semisphere also, due to the
shields which block rays coming from the southern part of the sphere (see
de~Maagt \etal 1998 for a more detailed discussion of their connection
with the optical configuration). 

The pattern has been calculated by using the Physical Theory of
Diffraction (PTD). Since pattern responses at levels smaller than
about  $-60$~dB are 
hard to measure, this is the most accurate method for predicting the
side lobe response on the antenna. 
The validity of the simulations at very small levels of the sidelobes
will be tested by measuring the antenna response of a fully 
representative copy of the {\sc Planck} 
telescope both in compact range and in outdoor far field
test range facilities.

\subsection{Simple estimates} 
\label{estimates}

Taking the level of Galactic emission and the antenna integrated response
from angular regions at different angles from the beam centre, we can provide
first order estimates of the expected GSC.

The region with $\theta_{bf}$ 
between $\simeq 1.2\deg$ and $\simeq 2\deg$ contains
about 0.5\% of the integrated response;
the region between $\simeq 1.2\deg$ and $\simeq 5\deg$ 
contains about 0.6\% of the integrated response
and all the rest of the far pattern ($\theta_{bf} \gsim 5\deg$)
contains about 1\% of the integrated response.
Of course, the remaining main integrated response falls
in the ``main'' beam (up to $\simeq 1.2\deg$) (see also Sect.~\ref{acc} for 
a discussion on the choice of these characteristic angles).
In addition, in the main spillover (2a) enters $\approx 0.1 \div 0.2 \%$ 
of the integrated response.

The sky signal at 30~GHz is known with pixel size
of about $2.5\deg$ by \COBE-DMR. 
For the present study at 30~GHz the relevant astrophysical source is the
Galaxy emission.
We have implemented ``small'' angular extrapolations
(see e.g. Burigana \etal 2000a for further details)
for generating Galaxy maps with resolution of about $1.2\deg$, corresponding
to Quad-Cube resolution 7.  

For simple estimates, we note that in the adopted 30~GHz Galaxy map
there are $\sim 13 \sq$ with a signal (in terms of antenna temperature $T_a$)
larger the 2~mK, $\sim 73 \sq$ with $T_a > 1.5$~mK and 
$\sim 230 \sq$ with $T_a > 1$~mK, while the minimum signal is $\sim 0.05$~mK
and about the 50\% of the sky shows a signal $\sim 0.1$~mK.

By combining these numbers with the percentages of integrated responses
falling within the above different angles from the beam centre, 
we expect to find
a contamination peaking at about 10$~\mu$K from the  pattern regions
between $\simeq 1.2\deg$ and $\simeq 5\deg$ and at few 
$~\mu$K from the  pattern regions
outside $\simeq 5\deg$.
In particular, in the main spillover (2a) we expect a signal peaking at
$\sim 2~\mu$K when it looks at high signal Galactic regions.
Similar contributions are expected from the pattern features
at few tens of degrees from the beam centre.
Of course, smaller contaminations ($\sim 0.5 \mu$K) are expected 
when the relevant pattern features look at regions with 
low signal Galactic.

Numerical calculations, like those presented 
in the next sections, are required for more accurate estimates.

\subsection{From data streams to sky maps}
\label{streamtomap}

The input map is converted from its original Quad-Cube pixelisation
to equal area, hierarchic HEALPix pixelisation scheme (G\'orski et al. 1998), 
adopted in the present work 
(see also section~\ref{acc} for details 
about the nominal resolution of this map). 
The final output of the simulation code relevant here are 
2 matrices with a number
of rows equal to the considered number of spin-axis positions $n_s$ 
for one year of mission ($n_s=180$ here) and 
a number of columns equal to the number of considered samplings 
along one scan circle ($n_p=180$ here). 
In the first matrix, ${\bf N}$, we store the pixel numbers corresponding to 
the main beam central
directions for the considered 180$\times$180 integrations; they are stored
in HEALPix pixelisation scheme 
at $n_{side}=32$ (the number of pixels $n_{pix}$ in a full sky map 
is related to $n_{side}$ by $n_{pix}=12 n_{side}^2$, G\'orski \etal 1998).
In the second matrix, ${\bf G}$, we store the antenna temperatures 
``observed'' by the considered
portion of the antenna pattern for the above pointing positions.
We neglect here the receiver noise and all the other systematics.
These data streams are the first output of our simulations; 
they give immediately
the impact of GSC 
and are useful to understand 
how this effect changes during the mission. 

From these data streams it is quite simple to obtain observed 
simulated maps, that
can be visualized for example in mollweide projection: 
we make use of ${\bf N}$ and ${\bf G}$ to simply coadd the temperatures
of those pixels observed several times during the mission. 
In this way we attribute to each pixel the average of the signals 
observed when the antenna pattern, due to the scanning strategy,
is differently oriented in the sky 
and Galactic regions with very different signal intensities enter in 
the intermediate/far sidelobes; 
of course, by coadding different samples of the same location into pixels
the systematic error per pixel is smaller than the systematic error in the
most contaminated sample.
This is because for different samples of the same location on the sky
the sidelobes are pointing towards different regions of the sky, some brighter
some fainter.

\subsection{Power spectra}
\label{streamtomap}

We can analyse both data streams and maps in terms of power at different 
scale-lengths or multipoles (or modes).

In order to analyse separately each scan circle of simulated data streams
we follow the approach suggested by Puget \& Delabrouille (1999) and decompose
the time series from the scan circle in Fourier series: 

\begin{equation}
g(\xi)=\sum_{m=0}^{m=n_p/2-1} a_m \sqrt{2} {\rm cos}(m\xi+\xi_{0,m}) \, ,
\end{equation}
 
\noindent
where $g(\xi)$ is the signal at the position 
identified by an angle $\xi$ (between 0 and $2 \pi$) 
on the considered scan circle. The $2\times n_p$ coefficients 
$a_m$ and $\xi_{0,m}$ can be easily computed from the time series 
by solving a linear system. The amplitude, $a_m$, of each mode $m$,
analogous to the multipole $\ell$ of a usual spherical harmonic 
expansion, has to be compared with that of the wanted 
signal and of other sources of noise. 
This approach is particularly interesting for 
the data analysis during the mission, as {\sc Planck} will continuously scan
different circles in the sky.

For analysing the maps of coadded signals we use the standard
approach of computing their angular power spectrum.
We produce maps in the HEALPix pixelisation scheme 
(G\'orski \etal 1998) which 
takes advantage from the isolatitude of the pixels 
for a quick generation of a map 
from the coefficients $a_{\ell m}$ of the spherical harmonic expansion
and vice versa (Muciaccia \etal 1997).
[We will show the angular power spectra in terms of
$\delta T_\ell (\nu) = \sqrt{\ell (2 \ell +1)C_\ell(\nu)/4\pi}$].
This is a very significant test for evaluating the GSC impact on 
{\sc Planck} science, as the estimation of angular 
power spectrum of CMB fluctuations
from the sky maps is one of the main objectives of {\sc Planck} mission.
It offers also the possibility
of directly comparing the GSC with 
astrophysical contaminations
and other sources of instrumental noise.
Of course, it produces a ``global'' estimate of GSC effect, 
useful in the analysis of the whole sky maps.

\section{Accuracy and tests}
\label{acc_test}

The antenna pattern is theoretically known from optical 
simulation codes at the desired accuracy and resolution compatible with 
the available computing time. The adopted grids have a resolution
much better than those
of currently available sky maps at {\sc Planck} frequencies, although
extrapolations both in frequency and in angular resolution 
of existing maps allow to produce more refined simulated maps
(e.g. Burigana \etal 1998, 2000a).
On the other hand, the knowledge of Galactic emission at high resolution
is not particularly relevant here and only small angular extrapolations
up to a resolution of about $1\deg$ 
has been implemented.
Given the currently available input map resolution and the adopted
simplified scanning strategies, we are able to derive the power
of GSC only up to $\ell \sim m \lsim 80$, a mode/multipole range 
satisfactory for the study of the smooth features in the 
intermediate and far pattern response. 
A very good agreement is found with the high resolution computations based
on faster Fourier expansion methods (Wandelt \& G\'orski 2000)
which will be practically necessary for whole sky convolutions 
at angular resolutions higher than those adopted in this work.
On the other hand, at larger multipoles 
the GSC effects are dominated by other systematics, as discussed in Sect.~5. 

\subsection{Sky, pattern grids and numerical accuracy}
\label{acc}

We are interested in producing signal data streams as they would be observed
separately by different angular regions of the antenna pattern
in order to understand the effect of the different 
pattern features as they project in the sky during the {\sc Planck} observations.
We have considered here three regions: $0\deg \le \theta_{bf} \le \theta_{bf,1}$
(main pattern); 
$\theta_{bf,1} \le \theta_{bf} \le \theta_{bf,2}$ (intermediate pattern); 
$\theta_{bf,2} \le \theta_{bf} \le 180\deg$ (far pattern).
We have choosen $\theta_{bf,1} = 1.2\deg$ and $\theta_{bf,2} = 5\deg$. 
Of course, the choice of $\theta_{bf,1}$ and $\theta_{bf,2}$ 
has to be appropriate to the considered antenna pattern:
for a given telescope design it depends mainly on the considered frequency
and only weakly on the exact feed location on the focal surface.
For the 30~GHz channel, the main beam can be accurately measured in flight
through planets (Mandolesi \etal 1998, Burigana \etal 2000b)
up to response levels of $\approx -30$~dB with respect to the peak response;
$\theta_{bf} \gsim 1.2\deg$ corresponds to antenna
responses lower than $-40$~dB, where
the beam response probably becames 
highly difficult to measure in flight;
$\theta_{bf,2} \sim 5\deg$ 
roughly divides pattern regions where significant response variations
occur on angular scales less than $1\deg$ from those where 
they occur on $\sim$ degree or much larger scales.

 
The observed antenna temperature is given by

\begin{equation}
T_{a,obs}= {\int J(\theta_{bf},\phi_{bf}) T_a(\theta_{bf},\phi_{bf}) d\Omega 
           \over \int J(\theta_{bf},\phi_{bf}) d\Omega } \, ,
\end{equation}

\noindent
where $J$ and $T_a$ are the antenna response and the sky antenna temperature
in the direction given by $\theta_{bf},\phi_{bf}$.

The convolution of antenna response with the sky and the integration of the antenna
pattern is simply computed by adding the contributions from 
all the pixels within the considered solid angle,
at resolutions corresponding to $n_{side}=1024, 512, 64$ 
respectively for the main, intermediate and far pattern 
in order to take accurately into account the pattern response variations. 
The main pattern is given in equispaced cartesian coordinates
with $\Delta x = \Delta y = 0.0005$.
The intermediate pattern and the far pattern  
are provided in equispaced ``GRASP8'' polar grids
with $\Delta \phi'_{bf} = 10\deg$ and 
$\Delta \theta'_{bf} = 0.1\deg$ and $0.5\deg$, respectively.
When we extract all the pixels in the sky that 
contribute to the convolution
within the considered solid angle, the exact central position 
of each pixel typically does not coincide 
with a grid point where the pattern is known. 
Simple standard bilinear interpolation (e.g. Press \etal 1992)
on the pattern grid has been implemented:
this is fast, robust and accurate enough for the present purposes. 

An estimate of the error introduced by 
the above discretizations and interpolation/computation methods 
can be provided by comparing the convolutions
obtained with different values of $n_{side}$,
for example by increasing $n_{side}$
to 1024 for the intermediate pattern convolution. The numerical error 
is negligible
($\lsim 1 \mu$K, $\lsim 0.5 \mu$K 
or $\lsim 0.2 \mu$K for the convolutions with the main, 
intermediate and far pattern, respectively).

\subsection{Tests with schematic skies and patterns}
\label{test}

Checking the consistency of the part of the code that
computes the signal entering in the main beam and in the intermediate
pattern is quite direct: we simply expect that the maps extracted from
the corresponding data streams are respectively very similar or 
roughly proportional 
(according to the fractional signal entering at 
$1.2\deg \le \theta_{bf} \le 5\deg$)
to Galactic emission pattern, except for the beam smoothing. 

Testing the validity of the computation of the signal entering 
the far pattern is not immediate, because it does not reflect
in a simple way the Galactic emission pattern.
We have verified the consistency of our simulation code 
by exploring simple cases for which we can 
easily predict the large scale symmetries of the maps derived from
the data streams observed by the far pattern. 
We have assumed a simple antenna pattern, centred on the optical axis
and perfectly symmetric
in $\phi_{bf}$, given by the sum of two gaussian shapes, one for the
main beam and one for the main spillover located at $90\deg$ from
the main beam centre, plus a constant low response level.
We have performed tests with the following different 
very simple input skies: 
$i)$ a spot at North Galactic pole:
it produces a map with a well defined slab on the Galactic plane;
$ii)$ a slab on the ecliptic plane: the corresponding map shows a 
signal maximum at the ecliptic poles and decreasing toward the ecliptic plane;
$iii)$ a slab on the Galactic plane: the corresponding map
shows a signal maximum at the Galactic poles and decreasing toward the Galactic
plane, where it exhibits small longitudinal modulations related 
to those of the solid angle subtended by the main spillover 
(this is due to the scanning symmetry with respect to ecliptic coordinates 
and not with respect to Galactic ones).
We have also verified that the angular power spectrum of the these maps 
presents a main peak at the multipole $\ell = 2$ and secondary peaks 
at its harmonic frequencies, as expected
from the $90\deg$ symmetry of the adopted far pattern. 

\section{Simulations results}
\label{results}

We have considered two options for the {\sc Planck} Carrier configuration.
The first is exactly that considered by de~Maagt \etal (1998) 
with $\alpha = 80\deg$ and including shields. In the second case 
we have used the same optical 
results but with $\alpha = 90\deg$: the 
corresponding results are then only
indicative, being not perfectly consistent because the spacecraft design
would be slightly different for this
configuration; on the other hand, 
this case is instructive because it allows to start addressing 
the question of the dependence of
GSC on a basic parameter of the scanning strategy.
We have considered the antenna pattern at 30~GHz 
with the beam centre located at $\theta_B=5.62\deg$ and $\phi_B=126.03\deg$.

\begin{figure*}[tb]
\vskip 1truecm
\centerline{
\psfig{figure=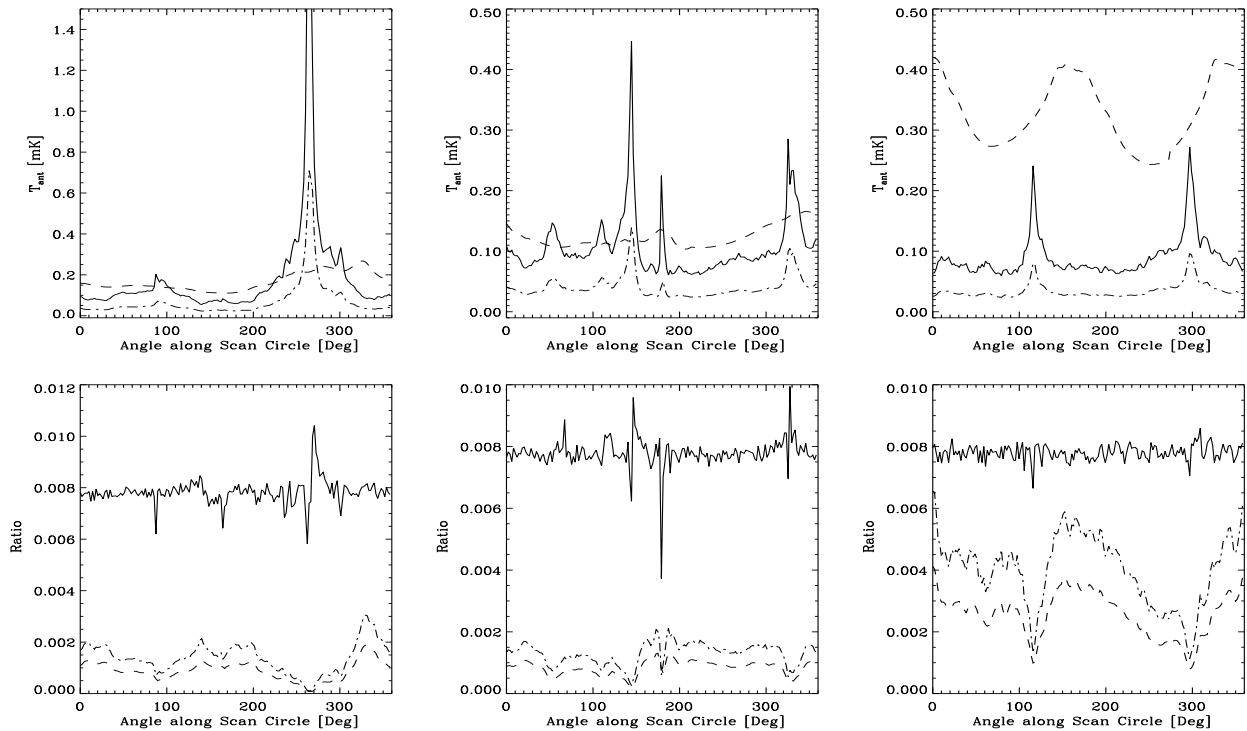,width=17.cm,height=10cm,angle=90}}
\caption[]{Top panels: absolute signals entering the main (solid lines), 
intermediate (multiplied by 50 for graphic purposes, dotted-dashed lines) 
and far (multiplied by 100, dashed lines) pattern 
for the case $\alpha = 80\deg$ for three scan circles corresponding 
to spin axis longitudes of 0\deg, 120\deg and 270\deg from left to right. 
The angle along the scan circle, ranging between 0\deg and 360\deg, 
is set to 0 at the beginning of the stream from each scan circle.
Bottom panels: the ratios between the signals shown in the top panels,
intermediate/main (solid lines), 
far/main (multiplied by 0.1, dotted-dashed line) and far/intermediate 
(multiplied by $5 \cdot 10^{-4}$, dashed lines).} 
\label{scans3_80}
\end{figure*}
\begin{figure*}[tb]
\vskip 1truecm
\centerline{
\psfig{figure=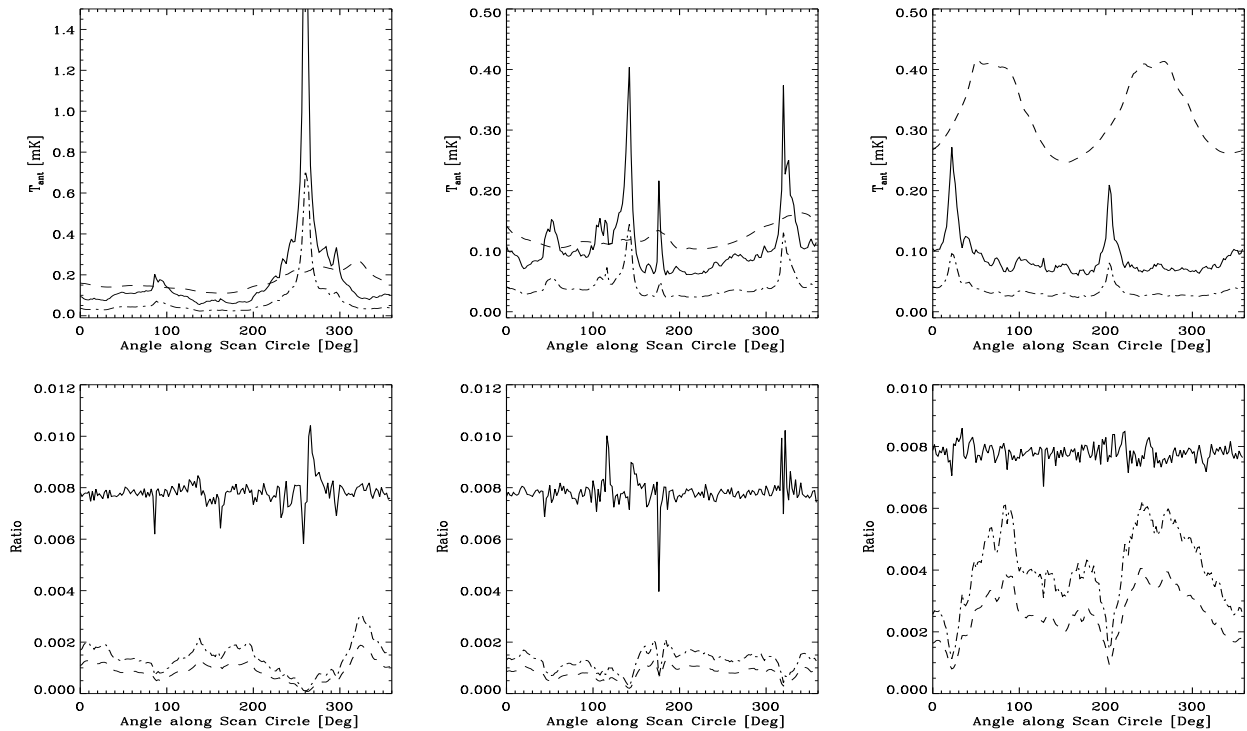,width=17.cm,height=10cm,angle=90}}
\caption[]{The same as in previous figure, but for the case $\alpha = 90\deg$.}
 \label{scans3_90}
\end{figure*}
\begin{figure*}[tb]
\vskip 1truecm
\centerline{
\psfig{figure=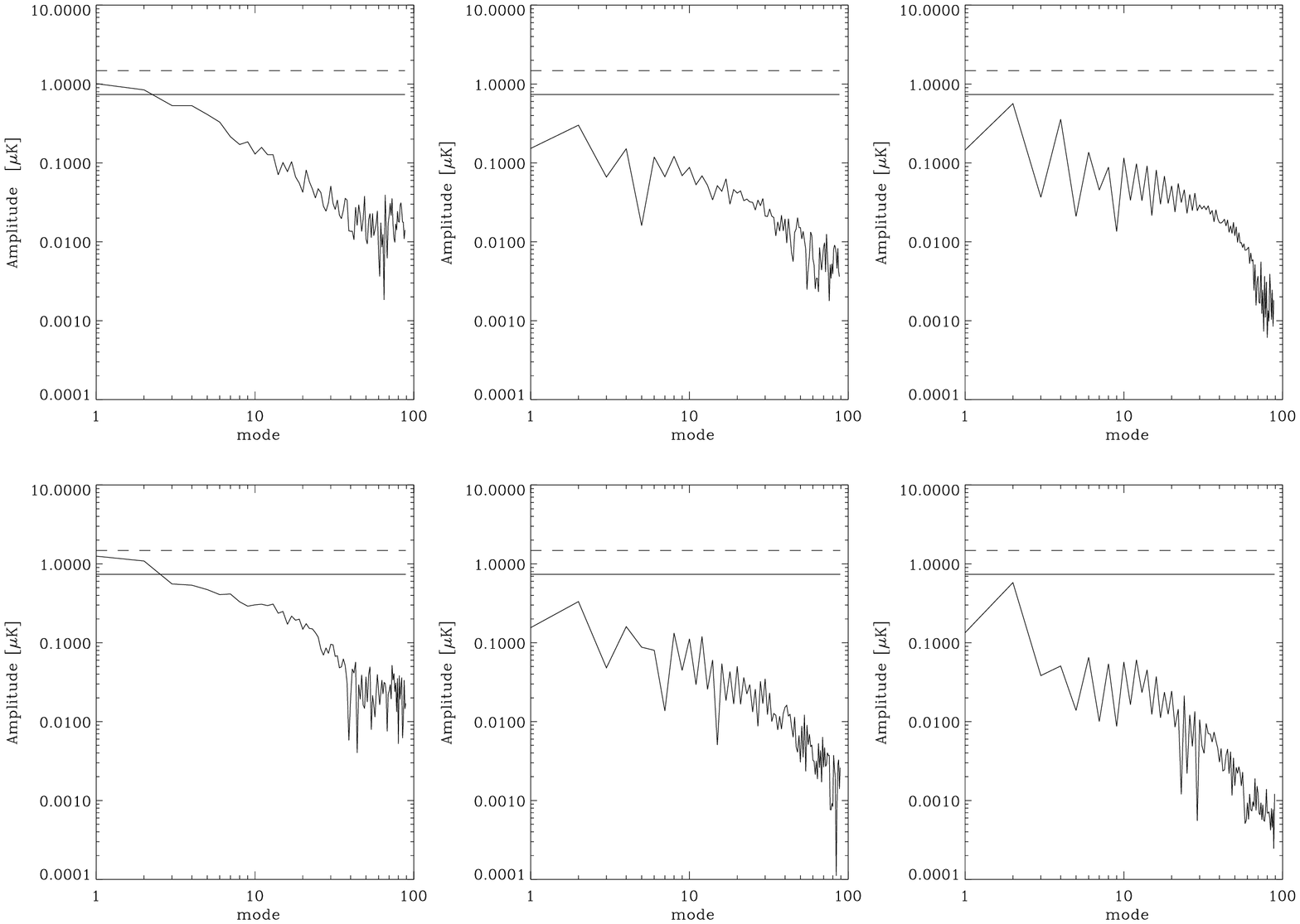,width=17.cm,height=10cm,angle=90}}
\caption[]{Fourier decomposition, see eq.~(1), 
of the signal at $\theta_{bf} \ge 1.2\deg$ from 
the scan circles of Fig.~\ref{scans3_80}  ($\alpha = 80\deg$, top panels) 
and \ref{scans3_90} ($\alpha = 90\deg$, bottom panels).
From the left to right: spin axis longitudes of 0\deg, 120\deg and 270\deg.
Dashed and solid horizontal lines represent the white noise
for one and four receivers, respectively. }
\label{fourier}
\end{figure*}

\subsection{Analysis of the scan circle data streams}

Figures~\ref{scans3_80} and \ref{scans3_90} show the absolute signals 
entering the main, intermediate and far pattern 
and their ratios for the data streams of three representative scan circles
for $\alpha = 80\deg$ and $90\deg$, respectively.

Note that the signal entering the intermediate pattern is
roughly proportional to that in the main beam:
two main relative maxima typically appear, related to the two crossings 
of the Galactic plane. 
The signal from the far sidelobes exhibits a clearly 
different and shifted angular behaviour, although
two main relative maxima are again typically 
present. These are mainly due to the contributions 
from the pattern features (1a) 
in the cases of the left and central panels,
and to the main spillover (2a) in the case of the right panels, 
as they cross the Galactic plane. Note in fact the displacement 
between the maximum signal from the main and far pattern, 
of few tens of degrees (several tens of degrees) 
for the left and central panel (right panel).

We have applied the Fourier series decomposition 
(see Fig.~\ref{fourier}) described by eq. (1)
to the sum of intermediate and far pattern (i.e. for $\theta_{bf} \ge 1.2\deg$)
data streams from the scan circles shown in Figures~\ref{scans3_80} and 
\ref{scans3_90}.

The same decomposition has been applied to white noise data streams, computed according
to the LFI sensitivity at 30~GHz (e.g. Maino \etal 1999) 
averaged over a number of scan circles that spans an ecliptic longitude length 
arc equal to the FWHM~$ = 33'$, i.e. essentially the sensitivity 
corresponding to half year mission.
The white noise power is above that of the signal entering at $\theta_{bf} \ge 1.2\deg$,
for practically all the modes $m \gsim 3$, becaming $\approx 10$ times larger 
at $m \approx 10$; this is essentially due to the strong decreasing 
of Galaxy fluctuations at small angular scales.

No significant differences are found by varying $\alpha$ from $80^{\circ}$
to $90^{\circ}$; for this reason and for sake of simplicity, 
we will show in what follows only the results for $\alpha = 80^{\circ}$, 
the angle for which the optical simulations have been
appropriately performed.

\begin{figure*}[tb]
\vskip 1truecm
\centerline{
\psfig{figure=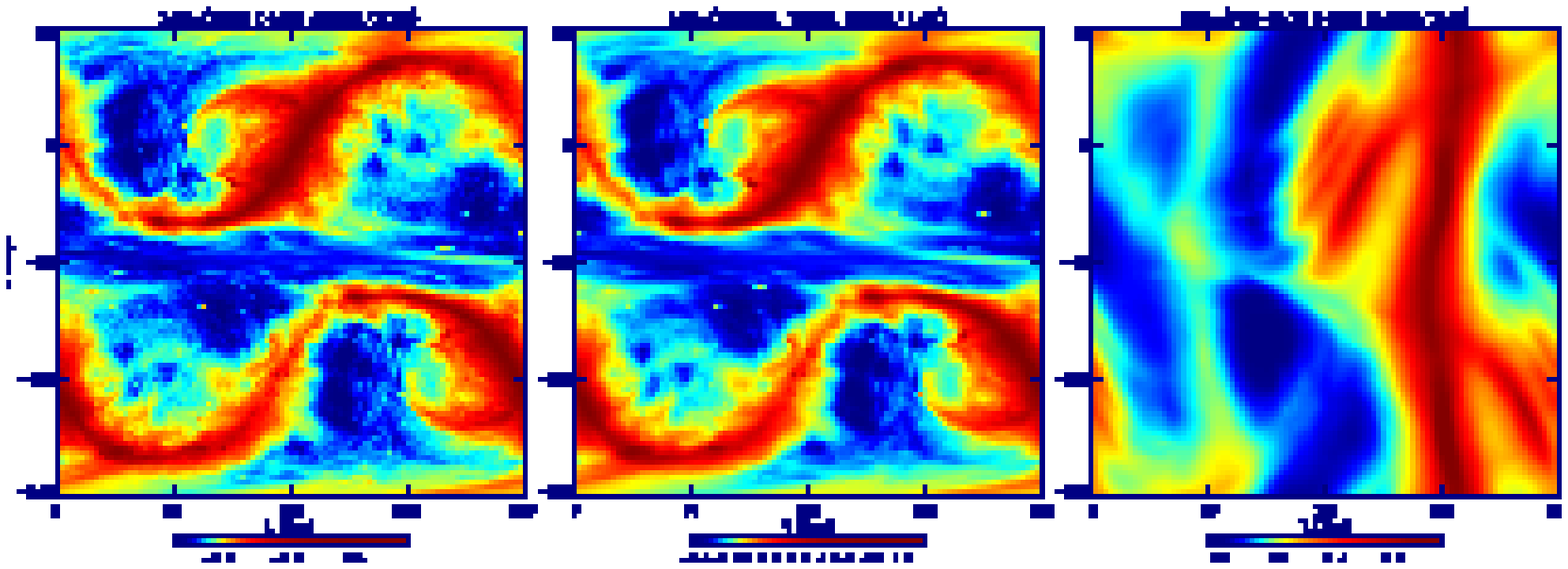,width=17.cm,height=10cm,angle=0}}
\caption[]{Synthetic view of the data stream from all scan 
circles for $\alpha = 80\deg$. The ecliptic coordinates $\lambda$ and $\beta$
properly refer here to the longitude and latitude of the telescope axis.
[For graphic purposes, in this plot the range 
of $\beta$ between $-80^{\circ}$ and $-240^{\circ}$ 
refer to the second half of each scan circle]. Compare with Fig.~4.}
\label{scansall_80}
\end{figure*}

All simulated data streams for a 1 yr mission 
are shown in Fig.~\ref{scansall_80}
for the case $\alpha = 80\deg$  (similar patterns are obtained in the case
$\alpha = 90\deg$). In the right panel, note the vertical high signal line 
at $\lambda \simeq 270^{\circ}$, corresponding to the main spillover (2a),
and the two high signal features, close to this line, at  
about $(\lambda , \beta ) \sim (180^{\circ},0^{\circ})$
and $(\lambda , \beta ) \sim (360^{\circ},-160^{\circ})$, corresponding to
the pattern features at few tens of degrees from the beam centre.
Note also how the azimuthally asymmetric 
far pattern reflects in the  
large difference between the two halfs (along $\lambda$ axis) of the right panel 
of Fig.~\ref{scansall_80} corresponding to the first and second six months
of observation.
This redundancy can be exploited for an efficient subtraction of GSC 
in the data analysis.

\subsection{Straylight contamination maps and angular power spectrum}

By coadding the data streams as described in Sect.~\ref{streamtomap}
we can obtain the corresponding maps.
This is shown in Fig.~8 for the case $\alpha = 80\deg$, 
by coadding the simulated data from the whole year.

\begin{figure}[tb]
\vskip 1truecm
\centerline{
\psfig{figure=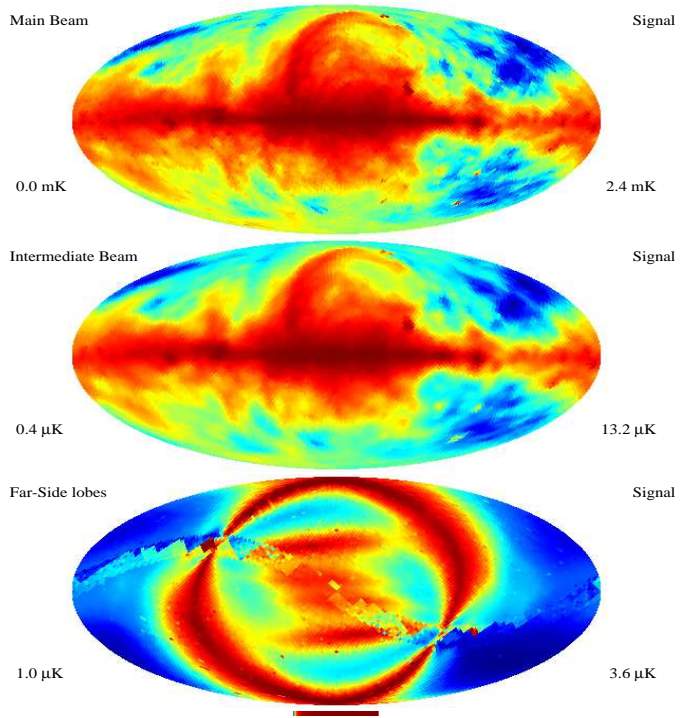,width=8.5cm,height=10cm,angle=0}}
\caption[]{Maps of GSC in Galactic coordinates from the coadding of the simulated data 
of the whole year for the case $\alpha = 80\deg$.
[For graphic purposes, the pixels close to the ecliptic poles, 
non observed because of the adopted
scanning strategy parameters, have been filled with values close to those
of the adiacient pixels]. In the bottom panel, 
note the two arcs corresponding to the main spillover (2a) and the central contaminations
corresponding to the pattern features at few tens of degrees from the beam centre.
They clearly correspond to the features in the right panel of Fig.~7.
Note also the wide ``stripe'' connecting the two ecliptic poles, which corresponds
to the sky region observed a single time in the current 1~yr simulation, owing to
the considered effective scanning strategy.}

\label{mapgsin1_90}
\end{figure}

As apparent in Fig.~8, 
the map from intermediate pattern is roughly proportional to that derived
from the main pattern; the relative intensities are roughly scaled by the
fraction of integrated response entering the two portions of the antenna pattern.

On the contrary, the sky ``observed'' by the far pattern is very different.
The signal is higher close to the Galactic plane, because of the features
in the antenna pattern within $10^{\circ} - 20^{\circ}$ from the main beam, and 
at about $90^{\circ}$
from the Galactic plane, because of the signal entering the main spillover (2a). 

\begin{figure}[tb]
\vskip 1truecm
\centerline{
\psfig{figure=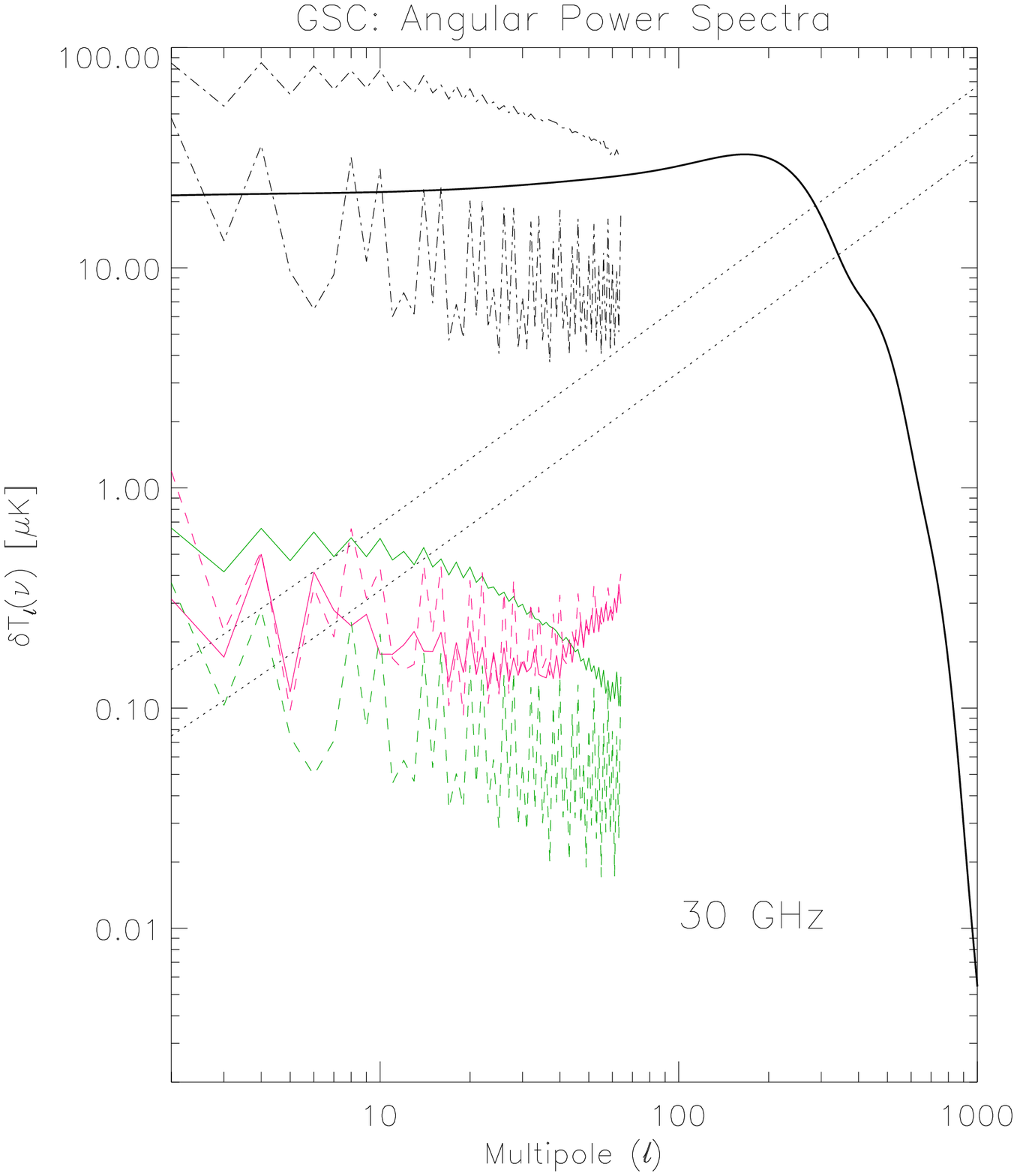,width=8.5cm,height=10cm,angle=0}}
\caption[]{Comparison between the angular power spectrum of GSC from different
pattern regions and that of the CMB anisotropy (thick solid line) 
and of the receiver white noise (dotted lines; upper line: a single receiver;
lower line: four receivers). 
Galaxy fluctuation power spectrum as seen by the main pattern
without map cuts (upper dotted-dashed line) and 
by considering only the regions 
at $\vert b \vert \ge 30\deg$ (lower dotted-dashed line).
Angular power spectra of GSC from the 
intermediate pattern without Galactic cut
(upper solid -- green -- line) and 
at $\vert b \vert \ge 30\deg$ (lower dashed -- green -- line) 
and from the 
far pattern without Galactic cut
(lower solid -- red -- line) and 
at $\vert b \vert \ge 30\deg$ (upper dashed -- red -- line).
See also the text. }
\label{comp_gsin_noise}
\end{figure}

We have computed the angular power spectra of these GSC maps 
(see Fig.~\ref{comp_gsin_noise}) 
and compared them with the theoretical angular power spectrum
of the white noise for a single 30~GHz receiver 
and for four receivers
and with a typical CMB anisotropy
angular power spectrum (a tilted - $n_p=0.9$ - power spectrum with standard
CDM cosmological parameters and approximately COBE normalized)
and with that of Galaxy fluctuations as seen by
the main pattern.
The most important contamination in terms of angular power spectrum derives
from the signal entering in the intermediate pattern when all the sky is considered;
on the contrary, if we consider only the regions at 
$\vert b \vert \ge 30\deg$, 
more crucial for {\sc Planck} main science,
the largest contributions to the GSC power spectrum 
derive from the far sidelobes. 
In general, the GSC power spectrum is larger than the white noise one
at low multipoles ($\ell \lsim 5$) but their ratio becames less than $\simeq 1/10$ 
at $\ell \simeq 50$ and decreases further at larger multipoles, 
due to their different dependence on $\ell$
\footnote{The increase of the power of GSC from far sidelobes 
at $\ell \gsim 50$, evident in the plot, is not an intrinsic effect of 
the GSC but is generated by the ``stripe'' corresponding to
the sky region observed a single time (see also Fig.~8) in the current 
1~yr simulation with $\alpha = 80^{\circ}$ and off-axis beam.}.

\section{Comparison with other sources of noise}
\label{comparison}

Many other sources of contamination, 
both instrumental and astrophysical in origin,
may affect {\sc Planck} observations.
We approach here a first comparison among the effects introduced by some of 
these systematics.

\subsection{Other kinds of instrumental noise}
\label{other_instr}

The impact of main beam distortions introduced by optical aberrations 
on {\sc Planck} measurements has been carefully
studied in several works 
(e.g. Burigana \etal 1998, 2000a, Mandolesi \etal 1997, 2000a).
Burigana \etal (1998) discussed the impact of the main beam distortions
for the representative case of an elliptical main beam shape.
In general, the absolute rms additional noise, in the range of few $\mu$K,
increases with the beam ellipticity.
The combined effect of main beam distortions
and of Galaxy emission fluctuations 
increases the additional error at $\sim 30$~GHz by a factor $\simeq 3$ 
with respect to the case of the essentially pure CMB fluctuation sky
at high Galactic latitudes, whereas it produces only a small additional 
effect at higher LFI frequencies. In addition, the combined effect
of main beam distortions and extragalactic source fluctuations is found
to be very small at all LFI frequencies 
(Burigana et al. 2000a) compared
to the noise induced by beam distortions in the case of a pure CMB sky.
Then, we focus here further on the impact of the main beam distortions
on the determination of angular power spectrum of CMB fluctuations
by considering the idealized case of the above pure CMB fluctuation sky.
The kind and the magnitude of optical distortions 
depend on the details of the optical design;
for aplanatic configurations currently under study (e.g. Villa \etal 1998 and
Mandolesi \etal 2000b) the typical main beam shape is roughly elliptical
owing to the strong reduction of the coma distortion. 
We computed a full year 
simplified simulation 
both for a pure symmetric gaussian beam 
with FWHM~$= \sigma \sqrt{8{\rm ln}2} =
30'$ and for an elliptical gaussian beam with axial 
ratio $r = 1.3$ and with the same effective resolution ($\sqrt{\sigma_x
\sigma_y} = \sigma$) of
the symmetric beam ($r=1$). We 
shift the spin axis at steps of 5$'$ and consider a step of 5$'$ between
two samplings on the same scan circle.
We computed the difference between the maps obtained from the elliptical and
the symmetric
beam by coadding the corresponding data streams and calculate the angular
power spectrum of this difference map
in order to understand which range of multipoles is
mainly affected.
As expected (see Fig.~10), this effect is particularly relevant 
at quite large multipoles, 
close to the CMB peak, where GSC power significantly decreases:
the magnitude, of course, is related to the value of $r$.

From optical simulations we know that $r$ typically increases with the distance
from the beam centre. From the present simulations we infer that a value
of $r$ quite smaller than 1.3, say less than $\simeq 1.1$ up to $-3$~dB from
the centre and less than $\simeq 1.2$ up to $-20$~dB from the centre, is good
enough for avoiding significant contaminations in the data, 
in agreement with the indications inferred
on the basis of the approximations of Burigana \etal (1998)
for the rms noise added by a main beam elliptical distortion.

\begin{figure}[tb]
\vskip 1truecm
\centerline{
\psfig{figure=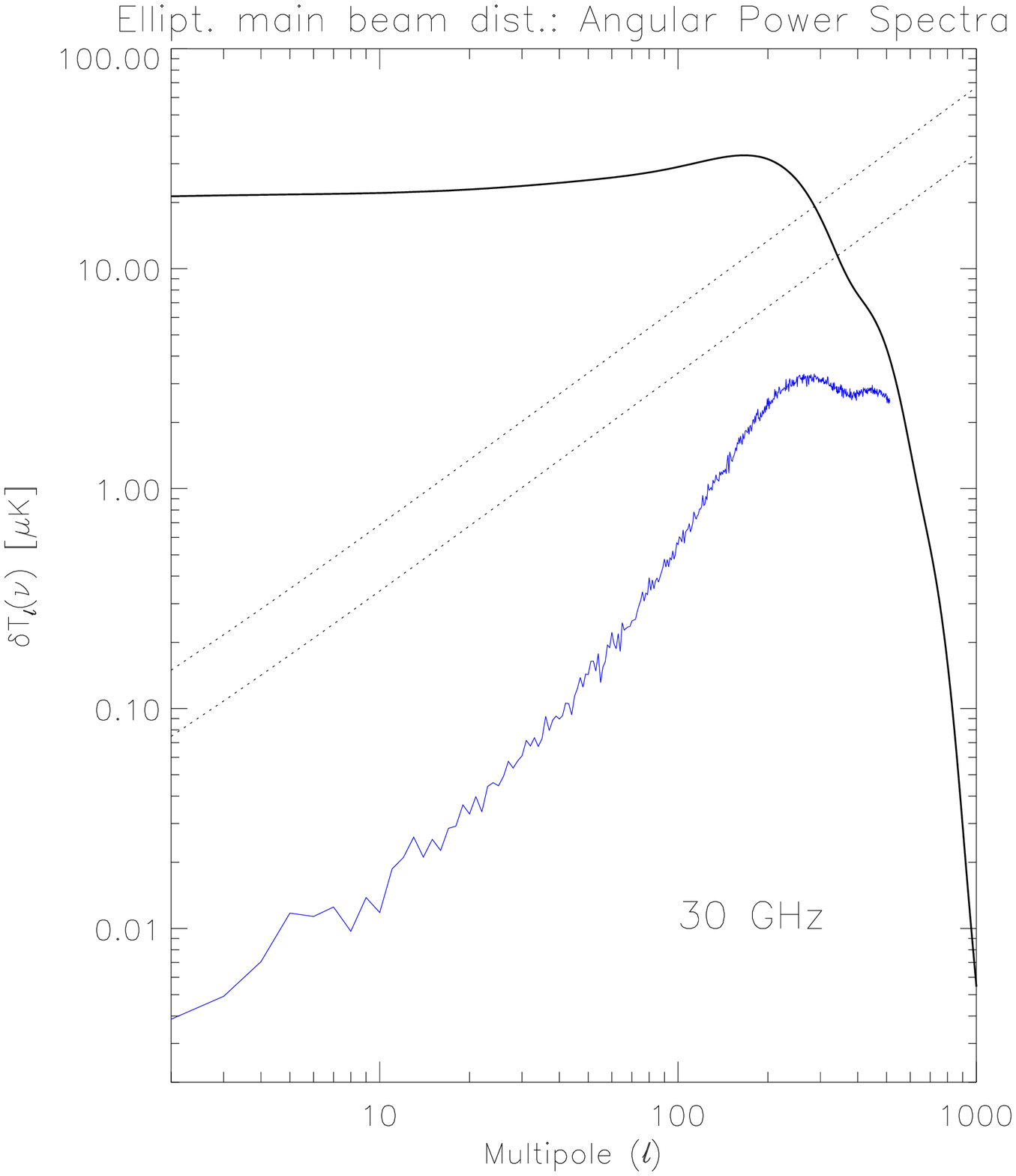,width=8.5cm,height=10cm,angle=0}}
\caption[]{Angular power spectrum introduced by an elliptical main beam distortion
with $r=1.3$ (lower -- blue -- line). See also the text.
}
\label{comp_instr_noise}
\end{figure}

The $1/f$ noise due to amplifier noise temperature fluctuations
induced by gain fluctuations in {\sc Planck} LFI receiver and its dependence
on the relevant instrumental parameters has been studied 
by Seiffert \etal (1997). It introduces additional noise in 
{\sc Planck} observations which shows as stripes in final maps 
(Janssen \etal 1996)
owing to the particular {\sc Planck} scanning strategy.
We have recently carried out detailed studies 
(Maino \etal 1999 and reference therein) on its effect on {\sc Planck} LFI 
measurements and on the efficiency of destriping algorithms
based on the use of the crossings between different scan circles  
(Bersanelli \etal 1996, Delabrouille 1998)
for a wide set of {\sc Planck} scanning strategies.
We extend here their simulations by relaxing the hypothesis of symmetric beam
to study the impact of main beam distortion into the destriping algorithm.
In Fig.~\ref{comp_instr_noise} we show the angular power spectrum of the receiver noise 
before and after applying the destriping algorithm when we 
include also the above elliptical distortion, 
for a simple scanning strategy with $\alpha =90\deg$ and a beam location
at $\theta_B = 2.8\deg$, $\phi_B=45\deg$, a ``mean''
choice regarding to the destriping efficiency (Maino et al. 1999). 
We find that the destriping efficiency is not 
significantly affected by the additional uncertainty 
introduced by the ``systematic'' differences among 
the observed temperatures resulting from different orientations of the main beam 
at the crossing points of different scan circles. 

\begin{figure}[tb]
\vskip 1truecm
\centerline{
\psfig{figure=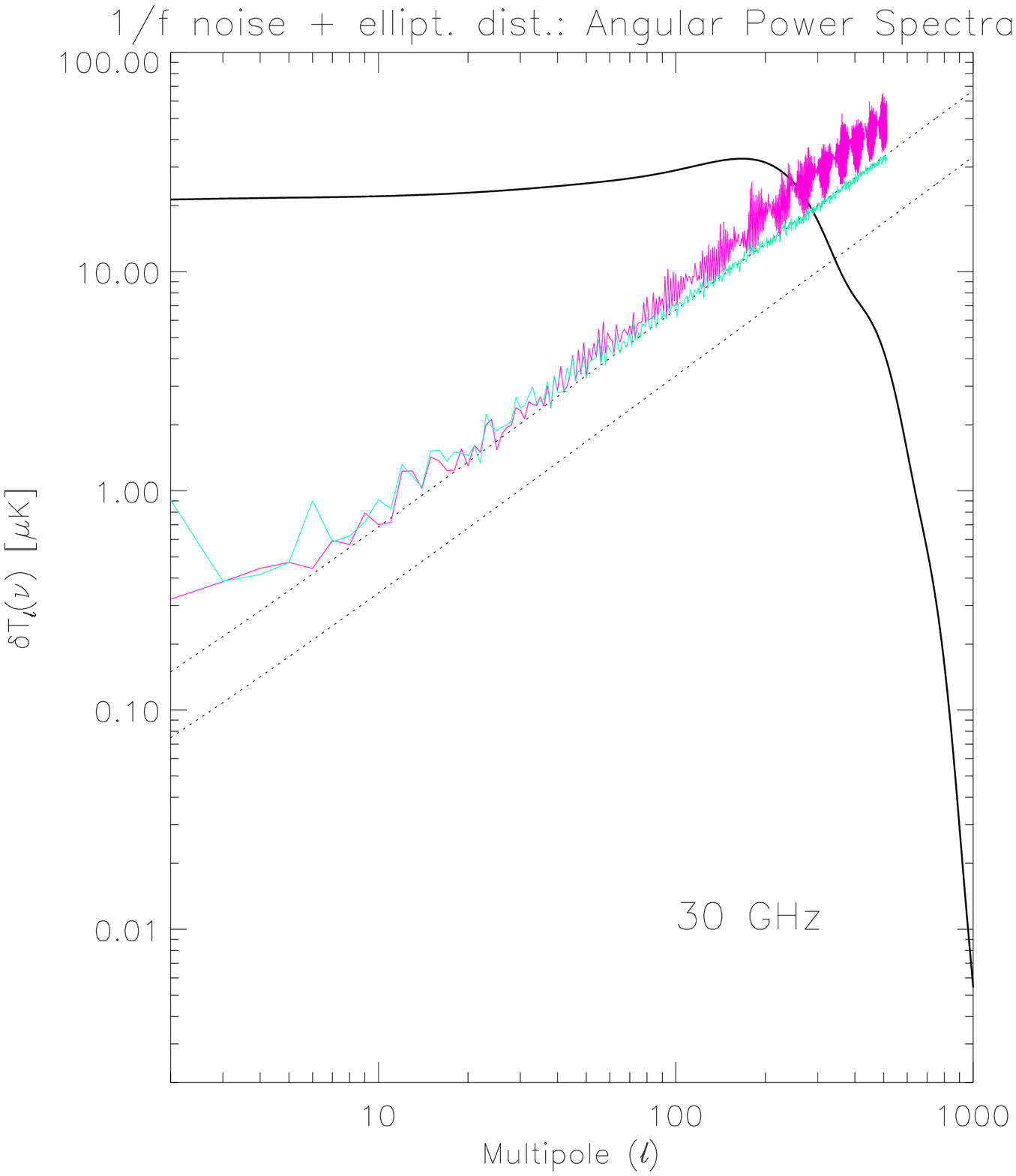,width=8.5cm,height=10cm,angle=0}}
\caption[]{Angular power spectrum of the $1/f$ noise before 
(upper -- red -- line with blobs) 
and after (lower -- green -- line, close to the level of theoretical white noise
power spectrum for a single receiver) the destriping
for the case of an elliptical main beam. }
\label{comp_instr_noise}
\end{figure}

As evident by comparing Figs.~$9\div11$, there is a crucial difference between
the angular power spectra of GSC, main beam distortion induced noise and $1/f$ noise.
The GSC affects particularly the determination of CMB angular
power spectrum at low multipoles, whereas main beam distortions
are critical at large multipoles. The $1/f$ noise affect both high 
and low multipoles, but destriping algorithms are particularly 
efficient in removing high multipole features in the power spectrum.
It is clear that all these effects have to be reduced both via
hardware and software. The $1/f$ noise can be reduced independently
of the other two, its magnitude being related essentially to the
instrument stability and to the scanning strategy (Maino et al. 1999).
On the contrary, a compromise has to be reached between GSC 
and main beam distortion noise, being both related mainly to the
optical design. As discussed in Mandolesi \etal (2000a), for
a given telescope design, their 
relative weight is controlled by the edgetaper.
The optical design has to be optimized to find a trade off 
for reducing the combined impact of these two effects.

\subsection{Astrophysical contamination}
\label{astro}

The impact of foregrounds on the primary cosmological goal of {\sc Planck} mission
has been extensively studied in literature for
what concern both Galactic and extragalactic contaminations,
of discrete and diffuse origin; 
{\sc Planck} itself will be a good opportunity for studying cluster 
physics, many classes
of extragalactic and Galactic sources and the diffuse emission
from the Galaxy (e.g. De~Zotti 1999a and references therein).
Many approaches have been studied to separate
the different components of the microwave sky and for deriving
their angular power spectra (e.g. Tegmark \& Esftathiou 1996, Hobson \etal 1998, 
Baccigalupi \etal 2000b, and references therein).

\begin{figure}[tb]
\vskip 1truecm
\centerline{
\psfig{figure=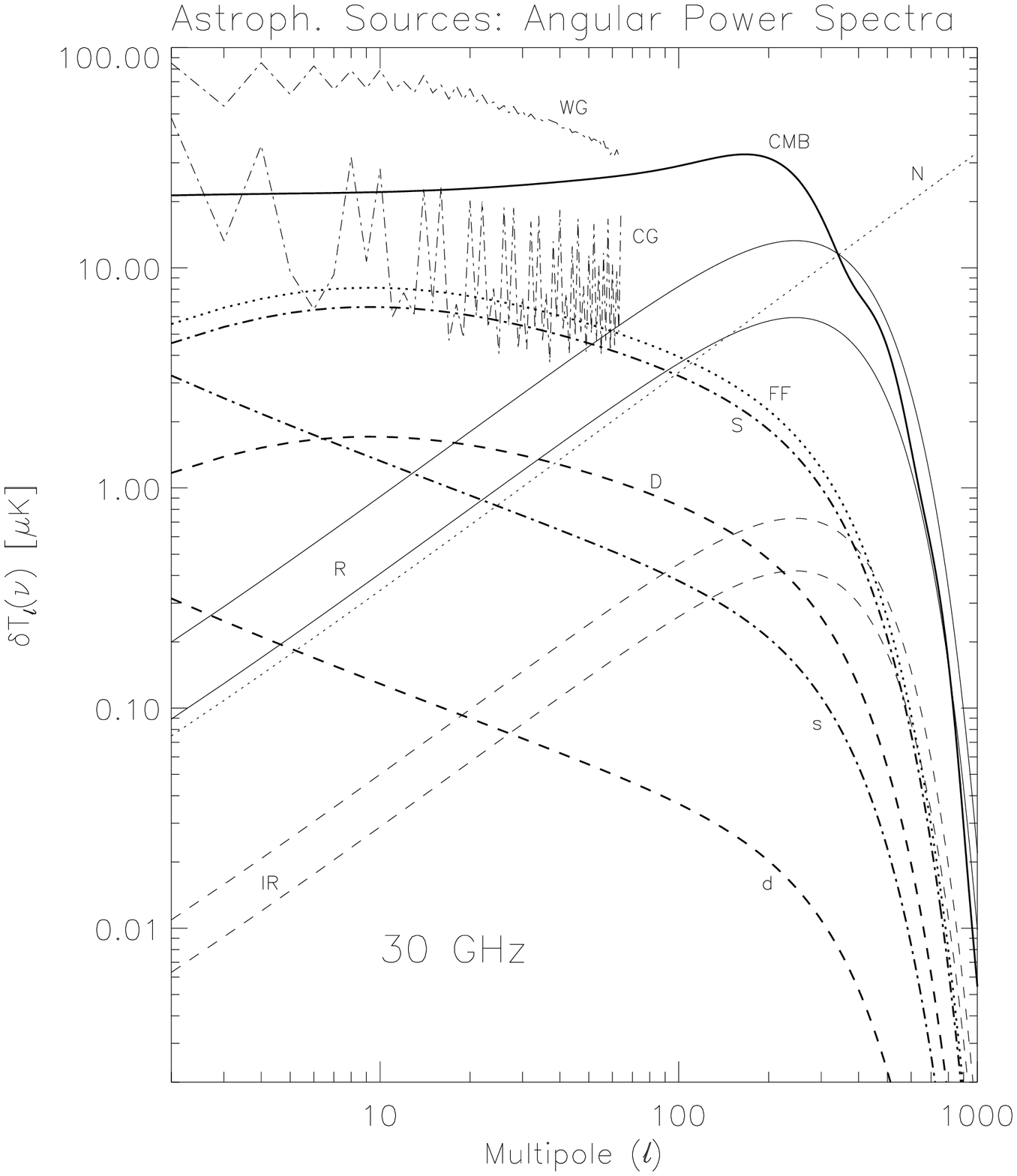,width=8.5cm,height=10cm,angle=0}}
\caption[]{Angular power spectrum of different sources of astrophysical contamination
compared with the CMB angular power spectrum (thick solid line, labelled with ``CMB'') 
and that of the white noise power spectrum for four receivers
(dotted line, ``N''). 
See the text for the meaning of the other lines and labels.}
\label{comp_astro_noise}
\end{figure}

We consider here the foreground 
impact on CMB science and their comparison with the effect of instrumental
systematics.
In Figure~\ref{comp_astro_noise} we report several estimates of the angular power 
spectra of different astrophysical components.

The Galaxy angular power spectrum is known to decrease with multipole order $\ell$:
we show here the power spectrum derived from the map observed by
the adopted main pattern  by cutting 
(lower thin dotted-dashed line, ``CG') or not (upper thin dotted-dashed line, ``WG'')
the region 
at $\vert b \vert \le 30\deg$ and the power spectra proposed by 
Tegmark \& Esftathiou (1996) for free-free (thick dotted line, ``FF''),
synchrotron (upper thick dotted-dashed line, ``S'') and dust (upper thick dashed line, ``D'') emission
at relevant Galactic latitudes.
We show also for comparison the power spectra for synchrotron 
(lower thick dotted-dashed line, ``s'') and dust (lower thick dashed line, ``d'') as derived 
by Prunet \etal (1998) and Bouchet \etal (1998) for
a sky patch at medium latitudes.
Of course, Galaxy contamination strongly depends on the considered region.

In terms of $\delta T_{\ell}$, Poisson fluctuations from extragalactic 
unresolved discrete sources increase approximately proportionally to 
the multipole $\ell$. 
We show here, separately for radiosource (thin solid lines, ``R'') and 
the far infrared galaxy (thin dashed lines, ``IR''),  
the Poisson fluctuation power spectra predicted by Toffolatti \etal (1998) 
as recently revised by Toffolatti \etal (1999), 
De~Zotti \etal (1999b) and references therein,  
on the basis of current source counts and assuming evolution 
models and spectra in agreement with current data, 
when sources above 1~Jy (upper curves)
or 100~mJy (lower curves) are detected and subtracted. Of course
radiosources dominate at low frequencies.
We have taken into account here a gaussian (FWHM=$33'$) beam smoothing 
in all cosmological and astrophysical angular 
power spectra and consequently neglected it in the receiver noise
angular power spectrum. 

At low multipoles, Galaxy contamination is larger then instrumental effects,
dominated by the GSC (and possible residual $1/f$ noise);
on the other hand, if the Galaxy emission and anisotropy can be modelled 
at few percent accuracy and at high Galactic latitudes, instrumental
effects can became comparable with the residual Galaxy contamination.

If not appropriately taken into account in the data analysis, main beam distortions
may introduce at high multipoles an additional contamination
comparable with that introduced from radiosource fluctuations, after their 
subtraction at few hundreds mJy level.

\section{Conclusions}
\label{conclusions}

We have studied the impact of GSC on {\sc Planck} observations at 30~GHz,
by considering different and complementary evaluation approaches:
absolute and relative quantification of the impact on scan circle 
data streams, Fourier decomposition of scan circle signals, 
computation of maps of GSC and evaluation of their angular power 
spectra. These different methods allow us to focus on different
aspects of GSC. 

No relevant differences are found by varying the angle $\alpha$ 
between the spin axis and the telescope line of sight 
from $80^{\circ}$ to $90^{\circ}$. 

Our simulations show that 
the GSC peaks at values of about $15 \mu$K (a value comparable with the sensitivity per pixel),
mainly due to the signal entering at few degrees from the beam centre.
Such values are found in the regions quite close to the Galactic plane,
where in any case the ``direct'' (i.e. observed by the main beam) 
contamination from the Galaxy prevents an accurate knowledge
of CMB fluctuations, as it is next to impossible to remove
the Galactic signals to accuracy better than $\sim$~ \%.
These large contamination values, although 
critical for CMB anisotropy measurements near the Galactic plane, 
are not crucial for the 
determination of Galaxy emission, which is several order of magnitude
larger. 

By considering all the pixels in the sky, the typical values of GSC 
are less than the 50\% of the white noise sensitivity per pixel.

The most crucial contamination 
derives from the signal entering in the far pattern, in spite of its 
peak values, of about 4$\mu$K, smaller than those obtained for the 
intermediate pattern regions.
In fact, although this effect does not seem to be very large in amplitude
(indeed, being nominally subdominant to power spectrum of expected receiver noise)
it does dominate the GSC at medium and high Galactic latitudes, 
which are critical regions for the extraction of the best quality results 
on CMB anisotropy.
This could be also critical for 
the {\sc Planck} polarization measurements which will take advantage 
from the two patches close to the ecliptic poles 
where the sensitivity will be several times better than the average,
according to the scanning strategy and the feed array 
arrangement.

As expected on the basis of the behaviour of Galaxy emission angular power spectrum,
the GSC affects the determination of the CMB angular power spectrum 
mainly in the low multipole region and much less at large multipoles, 
particularly when compared with the other instrumental effects considered here, 
the main beam distortion and the $1/f$ noise.
The additional noise introduced by the main beam distortion
can be in principle subtracted in the data analysis, provided that the beam pattern
is accurately reconstructed.

Of course, a substantial improvement in the data analysis is necessary  
to jointly treat all the systematics, of instrumental and astrophysical origin.
From the telescope design point of view, the best optimization of the edge taper
requires a trade off between the main beam distortion 
and the GSC effects.

\begin{acknowledgements}
We acknowledge stimulating and helpful discussion with
J.~Delabrouille and J.L.~Puget; we gratefully thank 
P.~de~Maagt and J.~Tauber for having promptly provided us 
with their optical simulation results.

\end{acknowledgements}

\noindent
{\bf References}
\rref{Baccigalupi C. \etal, 2000a, A\&A, submitted, astro-ph/0009135}
\rref{Baccigalupi C. \etal, 2000b, MNRAS, in press, astro-ph/0002257}
\rref{Balbi A. \etal, 2000, ApJL, submitted, astro-ph/0005124}
\rref{Bennet C. \etal, 1996a, ApJ, 464, L1}
\rref{Bennet C. \etal, 1996b, Amer. Astro. Soc. Meet., 88.05}
\rref{Bersanelli M. \etal, 1996, ESA, COBRAS/SAMBA Report on the Phase A
Study, D/SCI(96)3}
\rref{Bouchet F.R., Prunet S., Sethi S.K., 1999, MNRAS, 302, 663}
\rref{Burigana C. \etal, 1997, Int. Rep. TeSRE/CNR 198/1997}
\rref{Burigana C. \etal, 1998, A\&AS, 130, 551}
\rref{Burigana C. \etal, 2000a, Astro. Lett. Comm., 37, 253}
\rref{Burigana C. \etal, 2000b, A\&A, to be submitted}
\rref{De~Bernardis P. \etal, 2000, Nature, 404, 955}
\rref{Delabrouille J., 1998, A\&AS, 127, 555}
\rref{De~Maagt P., Polegre A.M., Crone G., 1998, {\sc Planck} -- Straylight Evaluation
of the Carrier Configuration, Technical Report ESA, PT-TN-05967, 1/0}
\rref{De~Zotti G. \etal, 1999a,  Proceedings of the EC-TMR Conference
``3 K Cosmology'', Roma, Italy, 5-10 October 1998,
AIP Conference Proc. 476, Maiani L., Melchiorri F., Vittorio N.,
(Eds.), pg. 204, astro-ph/9902103}
\rref{De~Zotti G. \etal, 1999b, New Astronomy, 4, 481}
\rref{G\'orski K.M. \etal, 1996, ApJ, 464, L11}
\rref{G\'orski K.M., Hivon E., Wandelt B.D., 1998,
to appear in ``Proceedings of the MPA/ESO Conference on
     Evolution of Large-Scale Structure: from Recombination to
Garching'',  Banday A.J. \etal (Eds.), astro-ph/9812350 }
\rref{Hanany S. \etal, 2000, ApJL, submitted, astro-ph/0005123}
\rref{Hobson M.P. \etal, 1998, MNRAS, 300,1}
\rref{Jaffe A.H. \etal, 2000, PRL, submitted, astro-ph/0007333}
\rref{Janssen M. \etal, 1996, astro-ph/9602009}
\rref{Lange A.E. \etal, 2000, PRD, submitted, astro-ph/0005004}
\rref{Maino D. \etal, 1999, A\&AS, 140, 383}
\rref{Mandolesi N. \etal, 1997, Int. Rep. TeSRE/CNR 199/1997}
\rref{Mandolesi N. \etal, 1998, {\sc Planck} LFI, A Proposal Submitted to the
ESA}
\rref{Mandolesi N. \etal, 2000a, A\&AS, 145, 323}
\rref{Mandolesi N. \etal, 2000b, Astro. Lett. Comm., 37, 151}
\rref{Muciaccia P.F., Natoli P., Vittorio N., 1997, ApJ, 488, L63}
\rref{Prunet S. \etal, 1998, A\&A, 339, 187}
\rref{Press W.H. \etal, 1992, Numerical Recipes in Fortran, second edition, 
Cambridge University Press}
\rref{Puget J.L. \etal, 1998, HFI for the {\sc Planck} Mission, A Proposal
Submitted to the ESA}
\rref{Puget J.L., Delabrouille J., 1999,
``HFI Sidelobe Straylight Requirement Document'' -- 2nd Revision
-- March 16, 1999}
\rref{Seiffert M. \etal, 1997, Rev. Sci. Instrum, submitted}
\rref{Smoot G.F. \etal, 1992, ApJ 396, L1}
\rref{Tegmark M., Efstathiou G., 1996, MNRAS 281, 1297}
\rref{Toffolatti L. \etal, 1998, MNRAS 297, 117}
\rref{Toffolatti L. \etal, 1999, 
Review talk at the Workshop ``The Sloan Summit on Microwave Foregrounds'',
Princeton, NJ, 14-15 November 1998,
de Oliveira-Costa A., Tegmark M. (Eds.) (ASP, San Francisco, 1999),
in press, astro-ph/9902343}
\rref{Villa F., Mandolesi N., Burigana C., 1998, Int. Rep. TeSRE/CNR 221/1998}
\rref{Wandelt B.D., G\'orski K.M., 2000, astro-ph/0008227, see also
TAC report ``Planck Straylight Modelling'',
http://www.tac.dk/~wandelt/papers.html}
\end{document}